\documentclass[aps,pra,superscriptaddress,preprint,amsmath,amssymb,showpacs]{revtex4-2}
\usepackage[english]{babel}
\usepackage{amssymb,bm}
\usepackage{graphicx}
\usepackage{amsmath}
\usepackage{color}
\usepackage{comment}
\usepackage[colorlinks=true,citecolor=blue,urlcolor=blue]{hyperref}
\begin{document}
 \begin{titlepage}
\title{Manifestation of  $a_1(1260)$ meson in the process $e^+e^-\to \pi^+\pi^-\pi^0$}
\author{I.V. Obraztsov}\email{ivanqwicliv2@gmail.com}
\author{A.S.Rudenko}\email{a.s.rudenko@inp.nsk.su}
\author{A.I. Milstein}\email{a.i.milstein@inp.nsk.su}
\affiliation{Budker Institute of Nuclear Physics, 630090 Novosibirsk, Russia}
\affiliation{\textit{Novosibirsk State University, 630090, Novosibirsk, Russia}}

%\date{\today}

\begin{abstract}
The charge asymmetry in the process $e^+e^-\to \pi^+\pi^-\pi^0$  is studied taking into account a longitudinal polarization of electrons (positrons). The asymmetry arises due to interference of amplitudes corresponding to production of pions in C-odd and C-even states. It is a manifestation of  $a_1(1260)$ meson in the intermediate state. Polarization leads to additional correlations in the differential cross section, which simplifies the experimental study of the process. It is shown that the charge asymmetry can reach several percent.
\end{abstract}

\maketitle
 \end{titlepage}
\section{Introduction}
Theoretical and experimental investigation of pseudovector mesons is one of the ways to study strong interactions at distances that cannot be described in the framework of QCD due to technical difficulties. Therefore, various phenomenological models have to be used. Although a fairly large amount of experimental information on pseudovector mesons has been accumulated at present, their study is still far from complete. One of possibilities to study pseudovector mesons is to measure the cross sections of meson production in $e^+e^-$ annihilation. For example, in the process $e^+e^-\rightarrow 4\pi$, an interesting phenomenon of $a_1(1260)\pi$ dominance  in the intermediate state at not very high energies has been discovered \cite{CMD-2:1998gab}. Another interesting  process is pseudovector meson production in $e^+e^-$ annihilation via two virtual photons. The point is that, according to the Landau-Yang theorem \cite{Landau:48, Yang:50}, particles with spin one cannot decay into two real photons due to Bose statistics. Therefore, studying the production of a pseudovector meson by two virtual photons allows one to understand the properties of corresponding form factor (i.e., the internal structure of particles).

At present, production of an isoscalar pseudovector meson $f_1(1285)$ by two virtual photons \cite{f1exp2020} has been observed, and the experimental results agree with the theoretical predictions \cite{f1th2020}. As for the isovector pseudovector meson $a_1^0(1260)$, there is currently no experimental evidence of its production by two virtual photons. Therefore, the study of this process is of undoubted interest.

\begin{figure}[h!]
	\centering
	\includegraphics[width=0.45\linewidth]{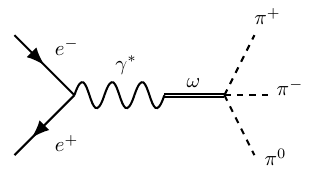}
	\includegraphics[width=0.45\linewidth]{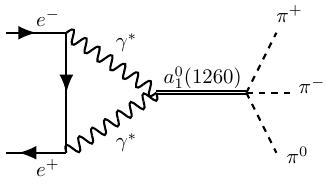}
	\caption{Diagrams corresponding to the process $e^+e^-\to\pi^+\pi^-\pi^0$.}
	\label{diagrams}
\end{figure}

 In our work, we study the contribution of $a_1^0(1260)$ meson, produced  by two virtual photons in the intermediate state, to the cross section of $e^+e^-\rightarrow \pi^+\pi^-\pi^0$ annihilation. This mechanism contributes to the amplitude of  $3\pi$ production in the state  with positive C parity. The main contribution to the cross section of $3\pi$ production is given by the amplitude $e^+e^-\rightarrow \gamma^*\rightarrow\omega\rightarrow \pi^+\pi^-\pi^0$, where  $3\pi$  have negative C parity. It is necessary to account for the  contributions of $\omega(782)$, $\omega(1420)$, $\omega(1650)$, and also $\phi(1020)$ mesons. Thus,  production of a virtual $a_1^0(1260)$ meson manifests itself in the charge asymmetry of the cross section due to interference of first and second diagrams in Fig.~\ref{diagrams}.

The process $e^+e^-\to\pi^+\pi^-\pi^0$ has been studied in many experiments (see, e.g., Ref.~\cite{EDSS2011} and references therein). However, in all these experiments the charge asymmetry was not measured. In our work we study the charge asymmetry taking into account the longitudinal polarization of an electron (positron) beam. The presence of such polarization can significantly simplify the study of various asymmetries in $e^+e^-$ annihilation \cite{Obraztsov:2023ljq}.

\section{Amplitudes $e^+e^-\rightarrow\omega\rightarrow 3\pi$ and $e^+e^-\rightarrow a_1\rightarrow 3\pi$}
In the center-of-mass frame, the amplitude $\mathcal{M}_{1}$, corresponding to the left diagram in Fig.~\ref{diagrams}, for electrons with helicity $\lambda$ and positrons with helicity $-\lambda$,  has the form
\begin{align}
	& \mathcal{M}_{1}=\dfrac{4\pi\alpha f_\omega \,\bm e_\lambda\cdot \bm J_\omega }{Q^2\,{\cal D}_\omega(Q)}\,,\nonumber \\
	&\bm e_\lambda=\bm e_x+i\lambda\bm e_y\,,\quad {\cal D}_\omega(q)=q^2-m_\omega^2+i\Gamma_\omega m_\omega\,.
\end{align}
Here $\alpha$ is the fine structure constant, $m_\omega$ and $\Gamma_\omega$ are the mass and width of $\omega$  meson, $\bm e_x$ and $\bm e_y$ are two unit vectors orthogonal to the electron momentum  $\bm P$ and to each other, $f_\omega$ is the coupling constant of virtual photon with $\omega$ meson, $Q=p_1+p_2+p_3=(E,0)$, $E$ is the total energy of  electron and positron in the center-of-mass frame; $p_1$, $p_2$, and $p_3$ are the momenta of $\pi^+$, $\pi^-$, and $\pi^0$, respectively. The vector $\bm J_\omega$ is determined from the amplitude ${M}_{\omega}$ of  transition $\omega\rightarrow 3\pi$, written in the covariant form,
\begin{align}\label{om3pi}
	& {M}_{\omega}=\dfrac{-2g_{\rho\pi\pi} g_{\omega\rho\pi}}{m_\omega}\left\langle \dfrac{1}{\mathcal{D}_\rho(p_1+p_2)}+\dfrac{1}{\mathcal{D}_\rho(p_1+p_3)}+\dfrac{1}{\mathcal{D}_\rho(p_2+p_3)} \right\rangle \varepsilon_{\mu\nu \alpha\beta}\omega^\mu p_1^\nu p_2^\alpha p_3^\beta\,.
\end{align}
Here $$ \mathcal D_\rho(q)= q^2-m_\rho^2+i\Gamma_\rho m_\rho\,,$$
$m_\rho$ and $\Gamma_\rho$ are the mass and width of $\rho$ meson, $g_{\rho\pi\pi}$ and $g_{\omega\rho\pi}$ are the corresponding coupling constants, $\omega_\mu$ is the polarization vector of $\omega$. From \eqref{om3pi} we find
\begin{align}\label{Jom}
	& \bm J_\omega=[\bm p_1\times\bm p_2]\,{F}_\omega \,,\nonumber\\
	& {F}_\omega=\dfrac{2g_{\rho\pi\pi}g_{\omega\rho\pi}\,E}{m_{\omega} }\left\langle \dfrac{1}{\mathcal{D}_\rho(p_1+p_2)}+\dfrac{1}{\mathcal{D}_\rho(p_1+p_3)}+\dfrac{1}{\mathcal{D}_\rho(p_2+p_3)} \right\rangle\,.
\end{align}
Finally,
\begin{align}\label{M1f}
	& \mathcal{M}_{1}=\dfrac{4\pi\alpha f_\omega F_\omega}{Q^2\,{\cal D}_\omega(Q)}\,\bm e_\lambda\cdot [\bm p_1\times\bm p_2]\,.
	\end{align}

Let us now  discuss the amplitude $\mathcal{M}_{2}$ corresponding to the right diagram in Fig.~\ref{diagrams}. We represent it in the form
\begin{align}\label{vertex_e+e-a1}
		& \mathcal{M}_{2}=\dfrac{1}{2}(4\pi\alpha)^2 f_\omega f_\rho\,\int \dfrac{d\bm k\,d\omega}{(2\pi)^4 k_1^2 k_2^2}\,{\cal J}^{ij}\,J_a^{ij}\,,\nonumber\\
	& k_1=(\omega,\bm k)\,,\quad k_2=(E-\omega, -\bm k) \,,
	\end{align}
where $f_\rho$ is the coupling constant of virtual photon with $\rho$ meson. The current ${\cal J}^{\mu\nu}$ corresponds to annihilation of $e^+e^-$ pair into two virtual photons with momenta $k_1$ and $k_2$. A straightforward calculation yields
\begin{align}
	& {\cal J}^{ij}=[ \omega e_\lambda^i N^j+(E-\omega) e_\lambda^j N^i] \,S^{(+)}
	-[e_\lambda^i k^j+e_\lambda^j k^i-(\bm k\cdot \bm e_\lambda)\delta^{ij}] \,S^{(-)}\,,\nonumber\\
	& S^{(+)}=\dfrac{1}{\mathcal{D}_{e}(P-k_1)}+\dfrac{1}{\mathcal{D}_{e}(P-k_2)}\,,\quad
S^{(-)}=\dfrac{1}{\mathcal{D}_{e}(P-k_1)}-\dfrac{1}{\mathcal{D}_{e}(P-k_2)}\,,
 \end{align}
where
$$ \bm N=\dfrac{\bm P}{|\bm P|}\,, \quad {\cal D}_e(q)=q^2-m_e^2+i0\,,$$ and $m_e$ is the electron mass.

The current ${J}_a^{\mu\nu}$ is determined from the matrix element ${M}_{\omega\rho a}$ of transition $\omega\rho\rightarrow a_1(Q)$,
\begin{align}
	{M}_{\omega\rho a}=ig_{a\omega\rho}\epsilon^{\mu\nu\alpha\beta}Q_\mu A^*_\nu\omega_\alpha \rho_\beta\,,
\end{align}
and from the matrix element ${M}_{a\rho\pi}$ of  transition $a_1(Q)\rightarrow \rho(p)\pi$,
\begin{align}
	{M}_{a\rho\pi}=-i\dfrac{g_{a\rho\pi}}{m_a}Q^\mu A^\nu[p_\mu\rho^*_\nu-p_\nu\rho^*_\mu]\,,
\end{align}
which determines the  matrix element of transition
 $a_1^0(Q)\rightarrow \pi^+(p_1)\pi^-(p_2)\pi^0(p_3)$,
\begin{align}
	& {M}_{a}=-\dfrac{2ig_{\rho\pi\pi}g_{a\rho\pi}}{m_a}A_\mu\,q_\nu \left[ \dfrac{p_1^\mu p_3^\nu-p_1^\nu p_3^\mu}{\mathcal{D}_\rho(p_1+p_3)}+\dfrac{p_2^\mu p_3^\nu-p_2^\nu p_3^\mu}{\mathcal{D}_\rho(p_2+p_3)}\right]\,.
	\label{vertex_a1_3pi}
\end{align}
Here $A_\mu$ and $m_a$ are the polarization vector and the mass of $a_1$, $g_{a\omega\rho}$ and $g_{a\rho\pi}$ are some constants given below, $\epsilon^{0123}=1$. As a result, we find the current $J_a^{ij}$,
\begin{align}
J_a^{ij}	& =-\dfrac{2iE^2g_{a\omega\rho}g_{a\rho\pi}g_{\rho\pi\pi}}{m_a\mathcal{D}_a(Q)} \left[ \dfrac{1}{\mathcal{D}_{\omega}(k_1)\mathcal{D}_{\rho}(k_2)}-\dfrac{1}{\mathcal{D}_{\omega}(k_2)\mathcal{D}_{\rho}(k_1)} \right]\nonumber\\
&\times	\epsilon^{ijl}\left[ \dfrac{\varepsilon_3p_1^l -\varepsilon_1 p_3^l}{\mathcal{D}_\rho(p_1+p_3)}+\dfrac{\varepsilon_3p_2^l -\varepsilon_2 p_3^l}{\mathcal{D}_\rho(p_2+p_3)}\right]\,.
\end{align}

Using the expressions for $\mathcal J^{ij}$ and $J_a^{ij}$, we obtain
\begin{align}\label{M2f}
	& \mathcal{M}_2= \lambda\bm e_\lambda\cdot[\bm p_1 Z(\varepsilon_1,\varepsilon_2)+\bm p_2 Z(\varepsilon_2,\varepsilon_1)]\,{F}_a\,,\\
		& {F}_a=-i\dfrac{2E\alpha^2g_{\rho\pi\pi}g_{a\rho\pi}g_{a\omega\rho}}{m_a \mathcal{D}_a(Q)}\,\mathcal{G}_a(E)\,,\nonumber\\
	& Z(\varepsilon_1,\varepsilon_2)=\dfrac{ E-\varepsilon_2}{\mathcal{D}_\rho(p_1+p_3)}+
	\dfrac{\varepsilon_2}{\mathcal{D}_\rho(p_2+p_3)}\,,\nonumber
\end{align}
where the function $\mathcal{G}_{a}(E)$ is defined by the integral
\begin{align}
 \mathcal{G}_{a}(E)	&=\dfrac{if_\omega f_\rho}{2\pi^2}\int\dfrac{d\bm k d\omega}{k_1^2 k_2^2}\,(k_1^2-k_2^2)\nonumber\\
	& \times\left[\dfrac{1}{\mathcal{D}_{\omega}(k_1)\mathcal{D}_{\rho}(k_2)}-\dfrac{1}{\mathcal{D}_{\omega}(k_2)\mathcal{D}_{\rho}(k_1)}\right] \left[\dfrac{1}{\mathcal{D}_{e}(P-k_1)}+\dfrac{1}{\mathcal{D}_{e}(P-k_2)}\right]\,.
\end{align}
\begin{figure}[h!] 
	\centering
	\includegraphics[width=0.6\linewidth]{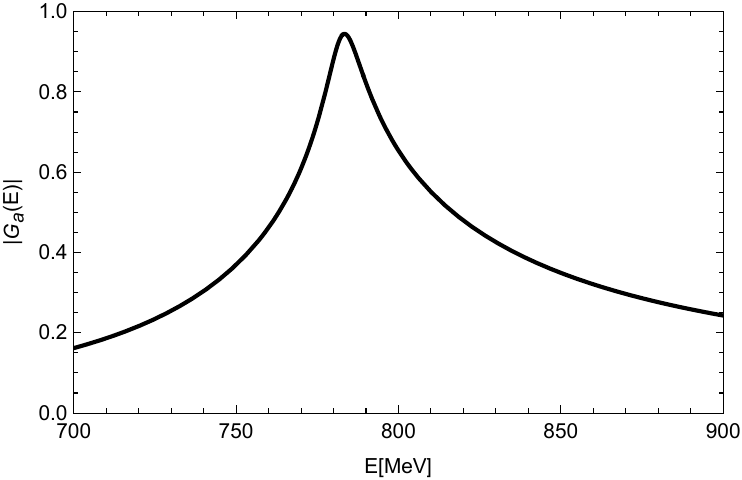}
	\caption{Dependence of dimensionless function $|\mathcal{G}_{a}(E)|$ on $E$ for $\omega(782)$ and $\rho(770)$.}
	\label{plot_FF_a}
\end{figure}
If we set $m_e=0$ in the function $\mathcal{D}_{e}(q)$, then the integral diverges logarithmically (collinear divergence). Carrying out calculations with logarithmic accuracy in $E/m_e$, we arrive at the following expression
\begin{align}	\label{FF_a1}
	& \mathcal{G}_a(E)=\dfrac{4f_\omega f_\rho}{E^2\mu_\omega^2\mu_\rho^2}\ln{\left(\dfrac{E}{m_e}\right)}\left[ \mu_\omega^2\ln{\left(1-\dfrac{E^2}{\mu_\omega^2}\right)}-\mu_\rho^2\ln{\left(1-\dfrac{E^2}{\mu_\rho^2}\right)} \right]\,,\nonumber\\
&\mu_\rho^2=m_\rho^2-i\Gamma_\rho m_\rho\,,\quad \mu_\omega^2= m_\omega^2-i\Gamma_\omega m_\omega\,. 
\end{align}
The dependence of dimensionless function $|\mathcal{G}_a(E)|$ on $E$ for $\omega(782)$ and $\rho(770)$ is shown in Fig.~\ref{plot_FF_a}. It is seen that  the function has a peak at $E\sim m_\omega,\,m_\rho$.

\section{Parameters of the model}\label{sec3}
\begin{figure}[h!]
	\centering
	\includegraphics[width=0.5\linewidth]{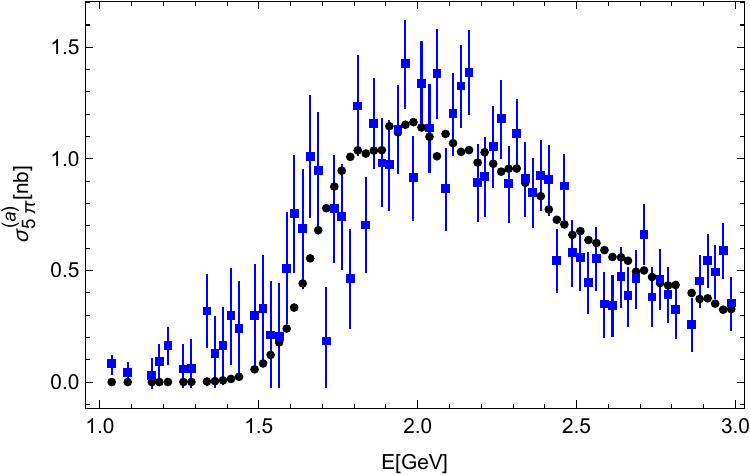}
	\caption{Comparison of experimental results (squares)  for $\sigma_{5\pi}^{(a)}=\sigma_{5\pi}^{(tot)}-\sigma_{5\pi}^{(\omega)}-\sigma_{5\pi}^{(\eta)}$ with the predictions of our model (dots) in the process $e^+e^-\rightarrow 2(\pi^+\pi^-)\pi^0$. Experimental data are taken from Ref.~\cite{BaBar:2007qju}. }
	\label{plots_2pi+pi-pi0}
\end{figure}

Let us now discuss the numerical values of parameters which we need to predict the cross section within our model. The parameters $f_\rho$, $f_\omega$, $g_{\rho\pi\pi}$, and $g_{\omega\rho\pi}$ for $\omega(782)$ are well known,
\begin{align}
	& \dfrac{f_\rho}{m_\rho^2}=0.2\,,\quad \dfrac{f_\omega}{m_\omega^2}=0.061\,,\quad g_{\rho\pi\pi}=5.94\,,\quad g_{\omega\rho\pi}=12.32\,.
\end{align}
For our purposes, we need to know the quantities $f_\omega g_{\omega\rho\pi}$ for $\omega(1420)$ and $\omega(1650)$, as well as $f_\phi g_{\phi\rho\pi}$. Using the experimental data \cite{BaBar:2004ytv,Achasov:2003ir} (see also \cite{SND:2020ajg}), we obtain 
\begin{align}
	&\dfrac{f_\omega g_{\omega\rho\pi}}{m_\omega^2}=-0.325\,[\omega(1420)]\,,\, 0.144 \,[\omega(1650)]\,,\quad \dfrac{f_\phi g_{\phi\rho\pi}}{m_\phi^2}=-0.063+i\,0.019\,. 
\end{align}	
We have estimated the constant $g_{a\rho\pi}$ from the decay width of $a_1$ \cite{Navas:2024ynf} under the assumption that the main contribution to this decay is given by the transition amplitude $a_1\rightarrow\rho\pi\rightarrow 3\pi$,
\begin{align}
	g_{a\rho\pi}=5.7\,.
\end{align}
The most difficult is to estimate the constant $g_{a\omega\rho}$ for $\omega(782)$. It is this constant which mainly determines the charge asymmetry in the energy range under consideration. We have found $g_{a\omega\rho}$ using the cross section $\sigma_{5\pi}^{(tot)}$ of the process $e^+e^-\rightarrow 2(\pi^+\pi^-)\pi^0$ \cite{BaBar:2007qju}.
The main contribution to $\sigma_{5\pi}^{(tot)}$ is given by the cross sections $\sigma_{5\pi}^{(\omega)}$ and $\sigma_{5\pi}^{(\eta)}$ of the processes  $e^+e^-\rightarrow\omega\pi^+\pi^-\rightarrow 2(\pi^+\pi^-)\pi^0$ and    $e^+e^-\rightarrow\eta\pi^+\pi^-\rightarrow 2(\pi^+\pi^-)\pi^0$, respectively. Assuming that the difference $\sigma_{5\pi}^{(a)}=\sigma_{5\pi}^{(tot)}-\sigma_{5\pi}^{(\omega)}-\sigma_{5\pi}^{(\eta)}$ is due to the mechanism
$e^+e^-\rightarrow a_1\rho\rightarrow 2(\pi^+\pi^-)\pi^0$, it is possible to estimate the parameter $g_{a\omega\rho}$. Since we fitted the experimental data in a fairly wide energy range $1\-- 3\,\mbox{GeV}$, in addition to the contribution of
$\omega(782)$ it is necessary to take into account $\omega(1420)$ and $\omega(1650)$ mesons, i.e., the amplitudes $e^+e^-\rightarrow \omega(1420)\rightarrow a_1\rho$ and  $e^+e^-\rightarrow \omega(1650)\rightarrow a_1\rho$. The best agreement between the predictions of our model and the experimental data is achieved for $|g_{a\omega\rho}|$, corresponding to $\omega(782)$ meson, in the following range
\begin{align}
	16\lesssim |g_{a\omega\rho}|\lesssim 20\,.
\end{align}
A comparison of experimental results \cite{BaBar:2007qju} for $\sigma_{5\pi}^{(a)}$ with the predictions of our model is shown in Fig.~\ref{plots_2pi+pi-pi0} for $g_{a\omega\rho}=18.5\,.$ Good agreement between theory and experiment is evident.
\begin{figure}[h!]
	\centering
	\includegraphics[width=0.5\linewidth]{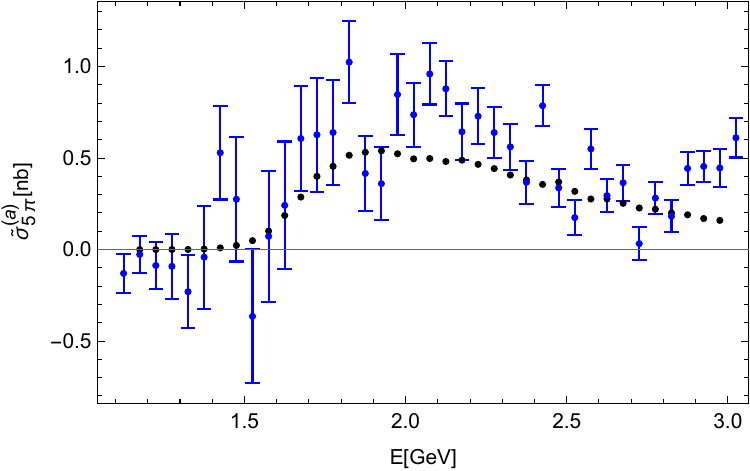}
	\caption{Comparison of experimental results (squares) in the process $e^+e^-\rightarrow \pi^+\pi^-3\pi^0$ for $\tilde{\sigma}_{5\pi}^{(a)}=\tilde{\sigma}_{5\pi}^{(tot)}-\tilde{\sigma}_{5\pi}^{(\omega)}-\tilde{\sigma}_{5\pi}^{(\eta)}$ with the results of our model (dots)  in the process $e^+e^-\rightarrow \pi^+\pi^-3\pi^0$. Experimental data are taken from Ref.~\cite{BaBar:2018rkc}.
	 }
	\label{plots_pi+pi-3pi0}
\end{figure}
Using the same values of parameters, we made a similar prediction for the contribution of $a_1$ to the cross section of the process $e^+e^-\rightarrow \pi^+\pi^-3\pi^0$ and compared it in Fig.~\ref{plots_pi+pi-3pi0} with the results of Ref.~\cite{BaBar:2018rkc}. As in the previous case, our predictions are consistent with the experiment.

\section{Charge asymmetry in the process $e^+e^-\to \pi^+\pi^-\pi^0$}
The differential cross section $d\sigma(\bm p_1,\bm p_2)$ for the process $e^+e^-\to \pi^+\pi^-\pi^0$ has the form
\begin{align}\label{dsig}
	& d\sigma(\bm p_1,\bm p_2)=\dfrac{\left|\mathcal{M}\right|^2}{16(2\pi)^5}\delta\left(B-2\bm p_1\cdot\bm p_2 \right)d\Omega_1d\Omega_2d\varepsilon_1d\varepsilon_2\,,\\
	& B=E^2-2E(\varepsilon_1+\varepsilon_2)+2\varepsilon_1\varepsilon_2+m_{\pi}^2\,,\quad \mathcal{M}=\mathcal{M}_1+\mathcal{M}_2.\nonumber
\end{align}
where $\mathcal{M}_1$ and $\mathcal{M}_2$ are given by Eqs.~ \eqref{M1f} and \eqref{M2f}, respectively, $m_\pi$ is the pion mass.  We assume that an electron has helicity $\lambda$, and the positron beam is not polarized.
The differential charge asymmetry is defined  as follows:
\begin{align}
 dA(\bm p_1,\bm p_2)	&=\dfrac{d\sigma(\bm p_1,\bm p_2)-d\sigma(\bm p_2,\bm p_1)}{2\sigma}\nonumber\\
	&=\dfrac{\mbox{Re} [\mathcal{M}_1^*\mathcal{M}_2]}{8(2\pi)^5\sigma}\delta\left(B-2\bm p_1\cdot\bm p_2 \right)d\Omega_1d\Omega_2d\varepsilon_1d\varepsilon_2\,,
\end{align}
where $\sigma$ is the total cross section of the process $e^+e^-\to \pi^+\pi^-\pi^0$. Note that the contribution of  $a_1$ meson to $\sigma$ can be neglected. Using the expressions obtained above, we find
\begin{align}\label{sig12}
	&\mbox{Re} [\mathcal{M}_1^*\mathcal{M}_2]=\Sigma_0+\lambda\Sigma_\lambda\,,\nonumber\\		
	&\Sigma_0=	\dfrac{2\pi\alpha|\bm p_1||\bm p_2|\,f_{\omega}}{E^2}\mbox{Im}\Bigg\{\dfrac{F_{\omega}^*F_a}{\mathcal{D}_{{\omega}}^*(Q)}\Bigg[
	(\bm N\cdot\bm p_1)\left[ B\, Z(\varepsilon_1,\varepsilon_2)+2\bm p_2^2\, Z(\varepsilon_2,\varepsilon_1) \right]\nonumber\\
	&- (\bm N\cdot\bm p_2)\left[B \,Z(\varepsilon_2,\varepsilon_1)+2\bm p_1^2\, Z(\varepsilon_1,\varepsilon_2) \right]\Bigg]\Bigg\}\,,\nonumber\\
	&\Sigma_\lambda=-\dfrac{4\pi\alpha|\bm p_1||\bm p_2|\,f_{\omega}}{E^2}(\bm N\cdot[\bm p_1\times\bm p_2])\nonumber\\
	&\times\mbox{Re}\Bigg\{\dfrac{F_{\omega}^*F_a}{\mathcal{D}_{{\omega}}^*(Q)}\left[(\bm N\cdot\bm p_1)Z(\varepsilon_1,\varepsilon_2)+(\bm N\cdot\bm p_2)Z(\varepsilon_2,\varepsilon_1)\right]\Bigg\}\,.
\end{align}
Recall that $\bm N$ is a unit vector parallel to the electron momentum. The asymmetry $dA$, integrated over all angles of the vectors $\bm p_1$ and $\bm p_2$, vanishes. Therefore, for the integral characteristic of the asymmetry, it is convenient to introduce the quantities
\begin{align}
	& A_0=\int \Xi_0(\bm p_1,\bm p_2)\,dA(\bm p_1,\bm p_2)\,,\quad
	A_\lambda=\int \Xi_\lambda(\bm p_1,\bm p_2)\,dA(\bm p_1,\bm p_2)
	\,,\nonumber\\
	&\Xi_0(\bm p_1,\bm p_2)= \theta(\bm N\cdot\bm p_1)\theta(-\bm N\cdot\bm p_2)-\theta(-\bm N\cdot\bm p_1)\theta(\bm N\cdot\bm p_2)\,,\nonumber\\
	&\Xi_\lambda(\bm p_1,\bm p_2)=\Big\{ \theta(\bm N\cdot[\bm p_1\times \bm p_2])-\theta(-\bm N\cdot[\bm p_1\times \bm p_2])\Big\}\nonumber\\
	&\times\Big\{\theta(\bm N\cdot\bm p_1)\theta(\bm N\cdot\bm p_2)-\theta(-\bm N\cdot\bm p_1)\theta(-\bm N\cdot\bm p_2)\Big\}\,.
\end{align}
Here $\theta(x)$ is the Heaviside theta function. It is important that $A_0$ is independent of the helicity  $\lambda$, and $A_\lambda$ is proportional to $\lambda$. Thus, the asymmetry $A_0$ is determined by the contribution of $\Sigma_0$ to the interference of amplitude, and $A_\lambda$ is determined by the contribution of $\Sigma_\lambda$, see \eqref{sig12}. Note that the function $\Xi_0(\bm p_1,\bm p_2)$ selects the region of momentum space in which $\pi^+$ and $\pi^-$ are in opposite hemispheres with respect to the vector $\bm N$, and the function $\Xi_\lambda(\bm p_1,\bm p_2)$ selects the region of momentum space in which $\pi^+$ and $\pi^-$ are in the same hemisphere. A sign of function $\Xi_\lambda(\bm p_1,\bm p_2)$ depends on that of the invariant $\bm N\cdot[\bm p_1\times \bm p_2]$.

\begin{figure}[h!]
	\centering
	\includegraphics[width=0.47\linewidth]{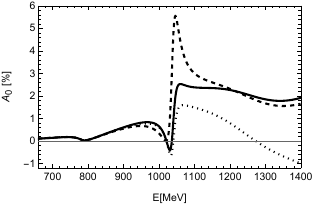}
	\includegraphics[width=0.47\linewidth]{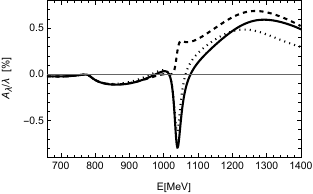}
	\caption{Dependence of $A_0$ (left) and $A_\lambda/\lambda$ (right) on $E$. Solid curves: contributions to the asymmetry of all $\omega$ mesons and $\phi(1020)$; dotted curves: contributions of only $\omega(782)$ and $\phi(1020)$; dashed curves: contribution of all $\omega$ mesons (excluding contribution of $\phi(1020)$). }
	\label{plots_asym_en}
\end{figure}

In Fig. \ref{plots_asym_en} the dependence of $A_0$ and $A_\lambda/\lambda$ on the energy $E$ is plotted for the parameters specified in section \ref{sec3} and $g_{a\omega\rho}=18.5\,$. It is seen that the asymmetry of $A_0$ can reach several percent, and the asymmetry of $A_\lambda$ is a few times smaller than $A_0$.

It is interesting to discuss which contributions to interference of amplitudes  are most significant for the dependence of asymmetry on the energy $E$. In Fig.~\ref{plots_asym_en}, the solid curves correspond to the contributions of all $\omega$ mesons and $\phi(1020)$, the dotted curves show  the contribution of only $\omega(782)$ and $\phi(1020)$, and the dashed curves show the contribution of all $\omega$ mesons (excluding the contribution of $\phi(1020)$). It is seen that  account for all mesons strongly affects the dependence of asymmetry on the energy. Therefore, measuring the asymmetry in the process $e^+e^-\to \pi^+\pi^-\pi^0$ is very important.

\section{Conclusion}
In our work,  the charge asymmetry in the process $e^+e^-\to \pi^+\pi^-\pi^0$ is studied with account for the longitudinal polarization of electrons (positrons). The asymmetry arises due to interference of amplitudes corresponding to production of pions in  C-even and C-odd states. The C-even state arises due to the contribution of  $a_1$ meson produced  by two virtual photons. The amplitude of latter process is determined by  $a_1\omega\rho$ interaction, which gives also one of  contributions to the cross sections $\sigma_{5\pi}^{(tot)}$ and $\tilde{\sigma}_{5\pi}^{(tot)}$ of  $e^+e^-\rightarrow 2(\pi^+\pi^-)\pi^0$ and $e^+e^-\rightarrow \pi^+\pi^-3\pi^0$ processes, respectively. It is shown that within  error bars the  experimental data on $\sigma_{5\pi}^{(tot)}$ and $\tilde{\sigma}_{5\pi}^{(tot)}$ are consistent with the assumption that they are a sum of $(2\pi)\omega$, $(2\pi)\eta$, and $\rho a_1$ contributions. Polarization of electrons (positrons) provides an additional opportunity to  study experimentally various correlations in the asymmetry. It is shown that the charge asymmetry can reach several percent, which makes its observation to be a difficult but realistic task.

\section*{Acknowledgement}
We are grateful to V.P.~Druzhinin and E.P.~Solodov for useful discussions. The work of I.V. Obraztsov was supported by the Foundation for the Advancement of Theoretical Physics and Mathematics "BASIS" under Grant No. 24-1-5-3-1.

\end{document}